\pdfoutput=1

\documentclass[11pt]{article}

\usepackage[]{acl}

\usepackage{times}
\usepackage{latexsym}

\usepackage[T1]{fontenc}

\usepackage[utf8]{inputenc}

\usepackage{microtype}

\usepackage{amsmath}
\DeclareMathOperator*{\argmax}{argmax} 
\usepackage{amssymb}
\usepackage{inconsolata}
\usepackage{graphicx}
\usepackage{booktabs}
\usepackage{enumitem}
\usepackage{bm}
\usepackage{multirow}
\usepackage{xcolor}
\usepackage{pifont}
\usepackage{tabularx}
\usepackage[export]{adjustbox}
\usepackage{array}
\usepackage{xspace}
\usepackage{subcaption}
\usepackage[normalem]{ulem} 
\usepackage{newunicodechar}
\newunicodechar{✓}{\ding{51}}
\newunicodechar{✗}{\ding{55}}

\title{Uncertainty Awareness of Large Language Models Under Code Distribution Shifts: A Benchmark Study}

\author{
Yufei Li$^{1}$
\xspace\xspace    
Simin Chen$^{2}$
\xspace\xspace    
Yanghong Guo$^{2}$
\xspace\xspace
Wei Yang$^{2}$
\xspace\xspace    
Yue Dong$^{1}$
\xspace\xspace    
Cong Liu$^{1}$\\
$^{1}$University of California, Riverside\xspace\xspace\xspace
$^{2}$University of Texas at Dallas\xspace\xspace \\
$^{1}$\texttt{\{yli927,yued,congl\}@ucr.edu}\xspace\xspace 
$^{2}$\texttt{\{sxc180080,yxg190031,wei.yang\}@utdallas.edu}
}

\begin{document}
\maketitle
\begin{abstract}

Large Language Models (LLMs) have been widely employed in programming language analysis to enhance human productivity.
Yet, their reliability can be compromised by various code distribution shifts, leading to inconsistent outputs.
While probabilistic methods are known to mitigate such impact through uncertainty calibration and estimation, their efficacy in the language domain remains underexplored compared to their application in image-based tasks.
In this work, we first introduce a large-scale benchmark dataset, incorporating three realistic patterns of code distribution shifts at varying intensities. 
Then we thoroughly investigate state-of-the-art probabilistic methods applied to CodeLlama using these shifted code snippets.
We observe that these methods generally improve the uncertainty awareness of CodeLlama, with increased calibration quality and higher uncertainty estimation~(UE) precision. 
However, our study further reveals varied performance dynamics across different criteria (e.g., calibration error vs misclassification detection) and trade-off between efficacy and efficiency, highlighting necessary methodological selection tailored to specific contexts.

\end{abstract}

\section{Introduction}

Large language models~(LLMs) have achieved impressive performance in code generation and analysis~\cite{codellama}. On the other side, current LLMs that are finetuned for specific tasks typically assume the test dataset is independently and identically distributed (i.i.d. or \emph{in-distribution}) with the training dataset~\cite{ovadia2019can}. This makes the reliability of such models for generalization a challenge, as real-world scenarios often come with \emph{distribution shifts} (also referred to as data drifts), such as codebase updates stemming from changes in library versions or new developers' contribution, leading to discrepancies in test dataset distribution and consequently, a degradation in the quality of generated code by the models.

\begin{figure}
    \centering
    \includegraphics[width=0.475\textwidth]{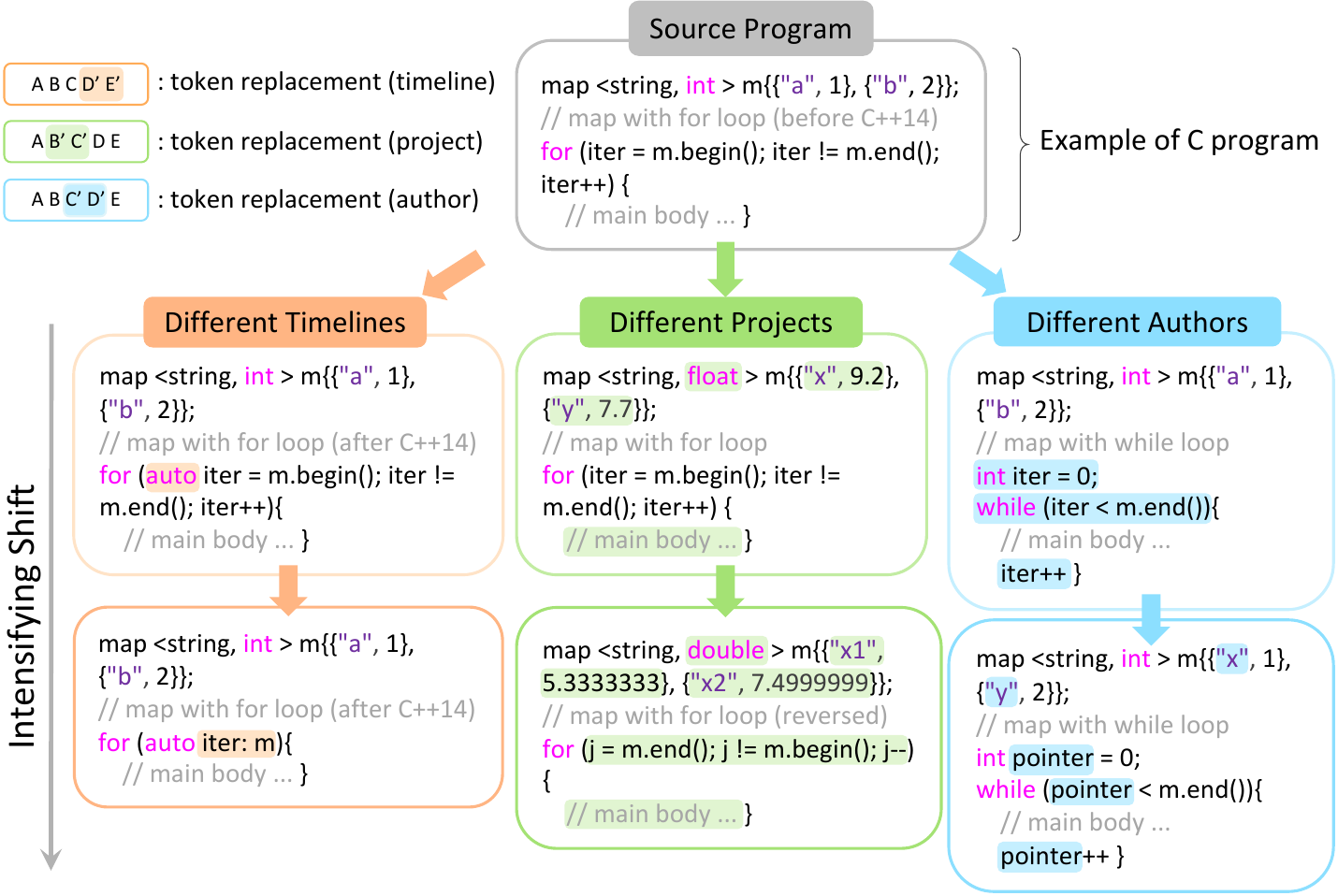}
    \caption{Three real-world code distribution shifts: \textsc{timeline shift}, \textsc{project shift}, and \textsc{author shift}, on C++ snippets of \emph{for} loops.}
    \label{fig:shifts}
    \vspace{-0.5cm}
\end{figure}

Addressing this challenge requires an understanding of the \emph{uncertainty awareness} of LLMs.
On the one hand, previous works show that deep models tend to be overconfident in their predictions despite low quality~\cite{zablotskaia-etal-2023-uncertainty,calibration}, particularly with shifted inputs.
This issue, known as miscalibration, highlights the disparity between a model's predictive confidence and its actual accuracy.
\emph{Uncertainty calibration}, therefore, emerges as a critical solution to improve prediction quality under distribution shifts.
Despite its significance, such notion has not received much attention in the code generation literature.
\emph{Uncertainty estimation}~(UE), however, is also vital even post-calibration, as mistakes are inevitable and it aids in assessing the reliability of model predictions, guiding decision-making on whether to accept or abstain from specific predictions~\cite{vazhentsev-etal-2022-uncertainty}.
UE methodologies include detecting error-prone instances, i.e., misclassification detection, and identifying out-of-distribution~(OOD) instances~\cite{ueDNN}, etc.

To investigate the uncertainty awareness of LLMs in the context of code distribution shifts, we define three prevalent shift patterns that cover most real-world evolution scenarios~\cite{codetimeevolve}: code changes due to library or API updates across \textsc{timeline shift}, code changes resulting from \textsc{project shift} that fulfill similar functions, code changes derived from \textsc{author shift}, as shown in Figure~\ref{fig:shifts}.
Unlike previous works in the language domain that only explore shift patterns~\cite{nie-etal-2022-impact,hu2023codes}, we further introduce a series of fine-grained shift intensities.
This allows for a more dynamic investigation of the performance of different methods under varying degree of shift intensities, as underscored by~\citet{ovadia2019can}.
We create a benchmark dataset by extracting and synthesizing Java code snippets from open-source projects, aligning them with each identified shift pattern and respective intensities.

Leveraging this dataset, we present the first comprehensive study of relative effectiveness of cutting-edge probabilistic methods in improving CodeLlama's prediction quality.
Specifically, we examine both classic approaches such as Monte Carlo Dropout~(MCD)~\cite{MCDropout}, Deep Ensemble~(DE)~\cite{DeepEnsemble}, and more recent techniques such as Mutation Testing (MT)~\cite{wang2019adversarial}, Dissector~\cite{wang2020dissector}.
Our findings reveal interesting dynamic patterns of efficacy.
For instance, adversarial method~\cite{wang2019adversarial}, though excels in identfying out-of-distribution~(OOD) examples, falls short in precisely predicting misclassifications; ensemble method is more robust against severe shifts compared to the post-hoc calibration~\cite{temperature}.
Additionally, we uncover a trade-off between calibration efficacy and efficiency for these methods, e.g., DE incurs a 50-fold increase in latency compared to deterministic baseline, underscoring the importance of selecting appropriate methods under specific task requirements.


Our contributions are:
\vspace{-0.3cm}
\begin{itemize}
    \item We develop a large-scale benchmark dataset consisting of three types of realistic code distribution shifts with varying intensities and OOD patterns.
    \vspace{-0.3cm}
    \item We adapt various probabilistic methods to the LLM setup and conduct an extensive study of their resultant uncertainty awareness under various shift patterns and intensities.
    \vspace{-0.3cm}
    \item We demonstrate that probabilistic methods generally mitigate the shift impact on LLM performance in code analysis, meanwhile revealing their performance variability and trade-off between efficacy and efficiency. 
\end{itemize}

\section{Related Work}

\noindent\textbf{Code Distribution Shifts} 
Recent studies have explored distribution shifts and the evaluation methods to assess the reliability of deep models.
For instance, semantic-preserving code transformations~\cite{rabin2021generalizability} present various shifts, ranging from common refactoring like variable renaming to more intrusive operations such as loop exchange.
\citet{nie-etal-2022-impact} evaluate the impacts of cross-project and time-segmented shifts on code summarization performance.
\citet{hu2023codes} assess the robustness of code analysis models under five shifts, including task, programmer, time-stamp, token, and concrete syntax tree.
These evaluation strategies resemble our defined shifts, e.g., time-segmented and time-stamp function similarly to \textsc{timeline shift}.   
However, some of these shifts, e.g., concrete syntax tree, are synthetic and unrealistic compared to our shift patterns that accommodate real-world evolution scenarios.
Also, our fine-grained shift scales allow a more thorough understanding of impacts of varying shift intensities.

\paragraph{Uncertainty Calibration}
Uncertainty is a natural aspect of predictive models, and modeling it properly can be  crucial for reliable decision-making. 
In recent years, research efforts have been directed towards quantifying uncertainty in deep learning (DL) models~\cite{lm_uncertainty_survey,vazhentsev-etal-2022-uncertainty}, which can be broadly classified into two categories: \emph{aleatoric} and \emph{epistemic}.
\citet{ue_ml} introduces several aleatoric uncertainty methods, such as Vanilla~\cite{hendrycks2016baseline} and Temperature Scaling~\cite{temperature}.
It also thoroughly explores epistemic uncertainty techniques, including both Bayesian Neural Networks~(BNNs) like Laplace approximation~\cite{mackay1992bayesian}, variational inference~\cite{blundell2015weight}, dropout-based methods~\cite{MCDropout, kingma2015variational}, and non-Bayesian methods such as ensembles~\cite{DeepEnsemble}, SWAG~\cite{maddox2019simple}, and SNGP~\cite{SNGP}.
This research lays the theoretical ground for our choice of uncertainty methods.
\citet{hu2023codes} alternatively applies uncertainty techniques, such as ODIN~\cite{liang2017enhancing} and Mahalanobis~\cite{lee2018simple}, to detect OOD examples. 
While it contributes significantly to the field, our study goes beyond identifying OOD datasets and offers a practical method for mitigating shift impacts via abstaining from low-quality predictions.

\section{Dataset Configuration}

\subsection{Three Real-world Code Shift Patterns}

Prior studies~\cite{zero-shot-llm} have demonstrated impressive zero-shot capability of LLMs on reasoning tasks, yet these models often fall short in complex tasks like code analysis (refer to Table~\ref{tab:zero_shot_F1} in Appendix~\ref{sec:additional_results}), leading to finetuning such models for improved performance necessary.  For example, 
consider a scenario where an LLM is finetuned on source-code of a project $P$. 
Over time, $P$ undergoes various changes such as file modifications and version updates, transforming to a new version $P'$.
Evaluating the model's performance on $P'$ is crucial to assess its reliability under distribution shifts across timelines~(\textsc{timeline shift}). 
Additionally, it's also important to determine if the model can efficiently adapt to different projects with similar functionalities (\textsc{project shift}), potentially saving training resources. 
Furthermore, code contributions from new developers, who possess distinct coding styles, introduce variability in coding patterns (\textsc{author shift}). These three real-world, commonly occurring code shifts define the focus of this project.

\begin{table}
    \centering
    \resizebox{0.48\textwidth}{!}{
    \begin{tabular}{ll|cccc}
    \toprule
    \multicolumn{2}{c|}{\textbf{Shift Pattern}} & \textbf{Train / Dev} & \textbf{Shift1} & \textbf{Shift2} & \textbf{Shift3} \\
    \midrule\midrule
    \multirow{5}{*}{\rotatebox[origin=c]{90}{\textbf{\textsc{Timeline}}}} & Snippet size & 12.66 / 12.64 & 12.58 & 12.62 & 12.71 \\
    & \# Snippets & 388,577 / 98,830 & 466,759 & 535,314 & 548,453 \\
    & Vocab & 17,858 & 17,874 & 17,849 & 17,862 \\
    & KL$\uparrow$ & \emph{0.0} & 0.15 & 0.26 & 0.32 \\
    & Cosine$\downarrow$ & \emph{0.0} & 0.96 & 0.94 & 0.91  \\
    \midrule
    \multirow{5}{*}{\rotatebox[origin=c]{90}{\textbf{\textsc{Project}}}} & Snippet size & 11.83 / 11.88 & 15.56 & 7.83 & 13.56 \\
    & \# Snippets & 172,591 / 41,216 & 106,607 & 50,279 & 116,714 \\
    & Vocab & 16,977 & 15,833 & 14,797 & 15,212 \\
    & KL$\uparrow$ & \emph{0.0} & 1.54 & 1.85 & 1.99 \\
    & Cosine$\downarrow$ & \emph{0.0} & 0.87 & 0.79 & 0.74 \\
    \midrule
    \multirow{5}{*}{\rotatebox[origin=c]{90}{\textbf{\textsc{Author}}}} & Snippet size & 16.08 / 15.28 & 15.85 & 14.61 & 15.32 \\
    & \# Snippets & 197,096 / 42,569 & 118,987 & 86,013 & 84,250 \\
    & Vocab & 16,118 & 16,180 & 16,834 & 16,566 \\
    & KL$\uparrow$ & \emph{0.0} & 0.12 & 0.35 & 0.66 \\
    & Cosine$\downarrow$ & \emph{0.0} & 0.91 & 0.87 & 0.81 \\
    \bottomrule
    \end{tabular}}
    \caption{Statistics of the three datasets with intensifying shifts (shift1 $\rightarrow$ shift3): \textsc{timeline shift}, \textsc{project shift}, and \textsc{author shift}. Snippet size is the average number of lines in each snippet. 
    }
    \vspace{-0.4cm}
    \label{tab:dataset}
\end{table}

\subsection{Benchmark Code Datasets}
To represent the three code shift patterns, we investigate seven open-source Java projects collected in the \emph{Java-small} benchmark\footnote{\url{https://s3.amazonaws.com/code2seq/datasets/java-small.tar.gz}}: \emph{elasticsearch}, \emph{gradle}, \emph{presto}, \emph{wildfly}, \emph{hadoop}, \emph{hibernate-orm}, and \emph{spring-framework}, which are language software for distributed computing and studied by existing code analysis literature~\cite{alon2019code2vec,code2seq}.
We extract Java files from each project as raw code snippets\footnote{We define a \emph{code snippet} as a single, complete function extracted from Java source code. Each function, including its signature and body, is treated as an independent unit of code.} and use Abstract Syntax Trees (ASTs) for tokenization.
To enable a more fine-grained study, for each shift pattern we create three levels of shift intensities (intensified from shift1 to shift3). 
The intensities are measured in two criteria: the Kullback–Leibler (KL) divergence between the token histogram distribution of each shifted set and dev set; the cosine similarity between embeddings encoded by CodeLlama of each shifted set and dev set.
The dataset statistics is shown in Table~\ref{tab:dataset}.
\vspace{-0.2cm}
\begin{figure}
    \centering
    \includegraphics[width=0.45\textwidth]{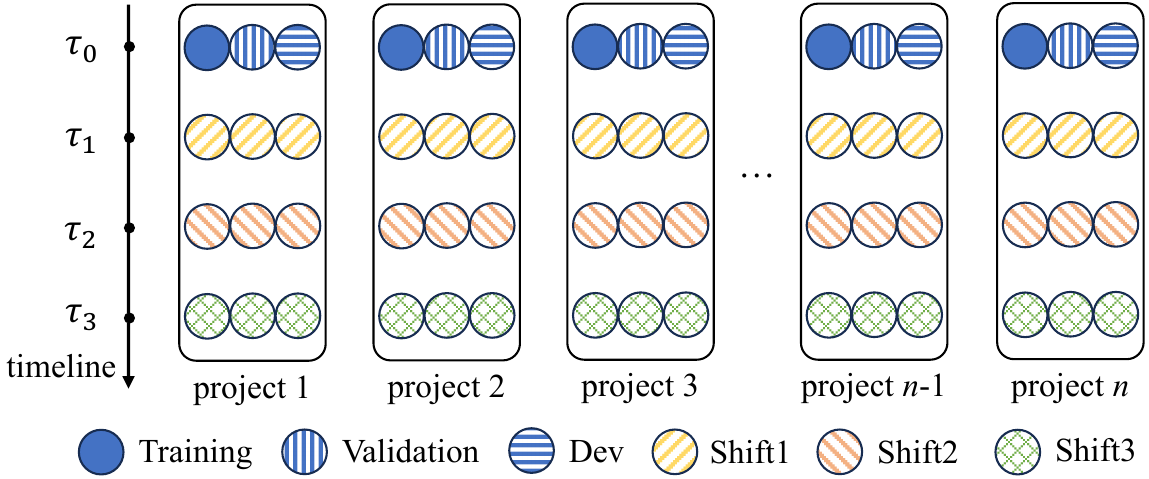}
    \caption{\textsc{Timeline shift} where all the $n$ projects are evaluated chronologically.}
    \label{fig:timeline_shift}
\end{figure}

\begin{figure}
    \centering
    \includegraphics[width=0.45\textwidth]{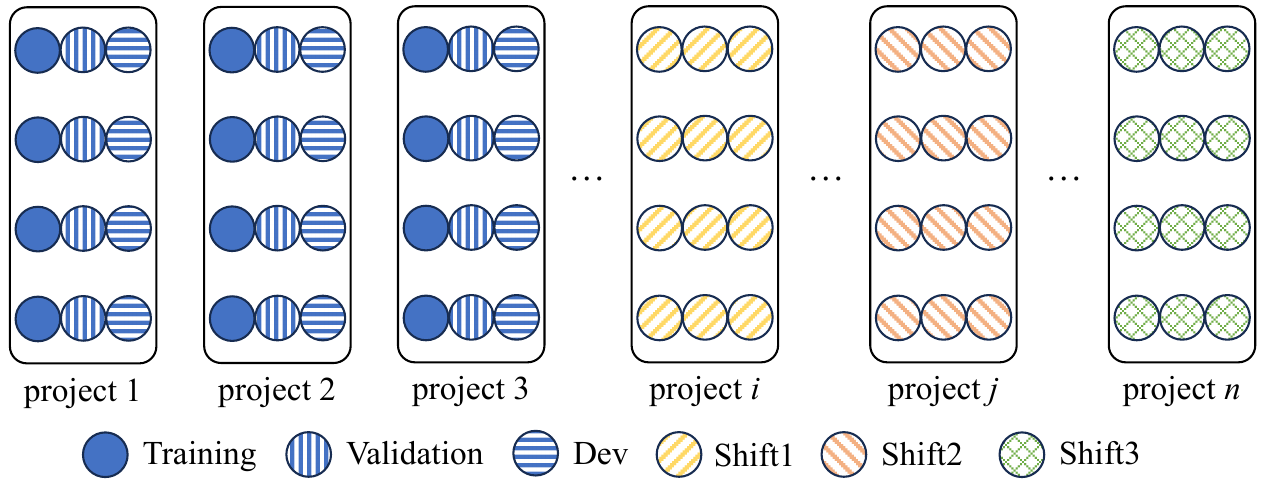}
    \caption{\textsc{Project shift} with cross-project split.}
    \label{fig:project_shift}
    \vspace{-0.3cm}
\end{figure}

\begin{figure}[t]
    \centering
    \includegraphics[width=0.45\textwidth]{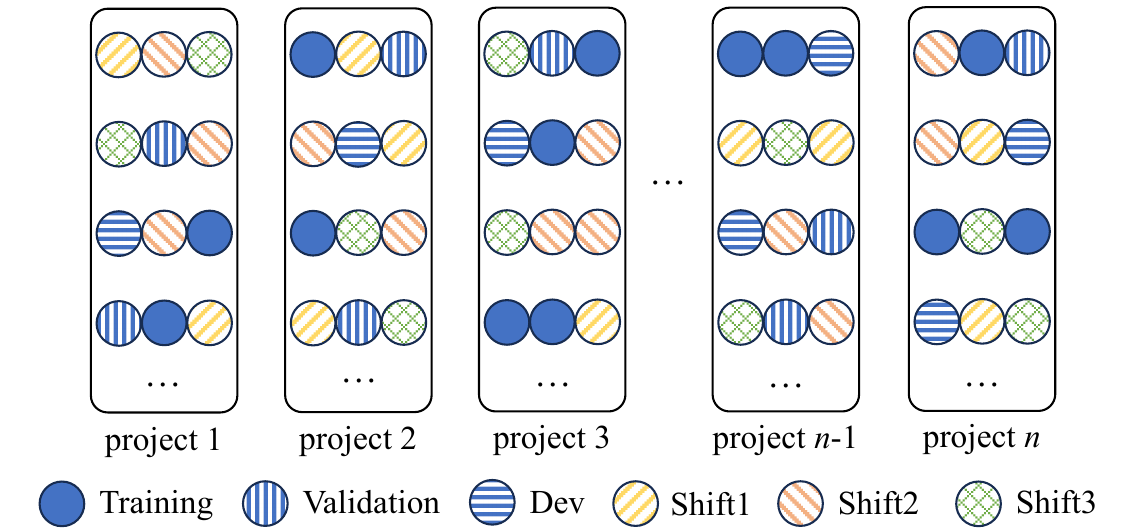}
    \caption{\textsc{Author shift} with cross-author split. Each color represents a unique author.}
    \label{fig:author_shift}
    \vspace{-0.3cm}
\end{figure}

\begin{itemize}[leftmargin=*]
    \item \textbf{Timeline Shift} As shown in Figure~\ref{fig:timeline_shift}, we consider four different release timelines, i.e., June 2020, December 2020, July 2021, January 2022, and collect Java files written by the same set of authors across times. Specifically, for each timeline, we check the commit history of all the projects and collect their corresponding release versions, e.g., the version of \emph{hadoop} around January 2022 is \textsc{release-3.3.2-RC1}.
    \item \textbf{Project Shift} As shown in Figure~\ref{fig:project_shift}, we calculate the KL divergence between each pair of the seven projects, and select top four projects, namely \emph{hibernate-orm}, \emph{presto}, \emph{spring-framework}, \emph{wildfly}, whose token distributions have least average divergence with others, as the training (and dev) set. The remaining three sets with increasing KL scores, namely \emph{hadoop}, \emph{gradle}, and \emph{elasticsearch} are noted as shift1, shift2, and shift3 datasets.
    \vspace{-0.3cm}
    \item \textbf{Author shift} As shown in Figure~\ref{fig:author_shift}, we split Java files into four groups. Although multiple programmers may have contributed to these files, we organize them based on the primary contributor, creating a semblance of \emph{author uniqueness} within each group. In this way, the \textsc{author shift} implicitly exists in our setup, as each group represents the work of a distinct set of programmers. Given the considerable size of the project and the presence of hundreds of contributors, we select four authors who have made the most significant number of commits throughout the project histories. The Java files committed by these authors are selected to form the datasets.
\end{itemize}

\section{Experiments}

\subsection{Experiment Setup}

Uncertainty methods have been applied to improve the reliability of LLMs. 
In this work, we focus on the state-of-the-art CodeLlama~\citep{codellama} (model implementation is detailed in Appendix~\ref{sec:model_implementation}).
We employ the following techniques to evaluate their effect on mitigating distribution shifts (method implementation and theory analysis are illustrated in Appendix~\ref{sec:uncertainty_setup}):
\begin{itemize}[leftmargin=*]
    \item \textbf{Vanilla Baseline} Our deterministic base model is CodeLlama-7B. The maximum softmax probability, also refer to winning score~(WS)~\citep{hendrycks2016baseline}, across the output space serves as the UE method. 
    \vspace{-0.3cm}
    \item \textbf{Temperature Scaling (TS)} \citet{temperature} propose a post-hoc calibration that learns a scalar parameter $T_{\text{ts}} > 0$ based on validation set $\mathbb{D}_{\text{val}}$ and align models' softmax probability more closely with the actual accuracy. TS ``softens'' the vanilla logit $l^c$ with $T_{\text{ts}}$ to obtain a new predictive distribution $p(y=c|x) = \frac{e^{l^c/T_{\text{ts}}}}{\sum_k{e^{l^k/T_{\text{ts}}}}}$. Since TS does not change the maximum of the softmax function, the class prediction remains \emph{unchanged}. We apply Shannon entropy of the new predictive distribution as the UE method.
    \vspace{-0.1cm}
    \item \textbf{Monte-Carlo Dropout (MCD)} \citet{MCDropout} introduce a Bayesian method that estimates \emph{inner-model} epistemic uncertainty using the Monte-Carlo average of $T_{\text{mcd}}$ dropout samples, which are generated using the same architecture but with different random seeds at dropout layers. We define calibrated outputs as the average logit over the $T_{\text{mcd}}$ dropout samples. Following~\citet{vazhentsev-etal-2022-uncertainty}, we consider three UE techniques: sampled winning score (SWS), probability variance (PV)~\cite{pmlr-v70-gal17a,uncertainty-adv-detection}, and Bayesian active learning by disagreement (BALD)~\cite{BALD}.
    \item \textbf{Deep Ensemble (DE)} \citet{DeepEnsemble} measure the \emph{cross-model} epistemic uncertainty by training $T_{\text{de}}$ independent vanilla models and averages all. We use the same model architecture but different initial seeds. We quantify uncertainty via the same UE methods as MCD.
    \item \textbf{Mutation Testing (MT)} \citet{wang2019adversarial} propose an adversarial method that measures the sensitivity of an input to model mutation operations, effectively quantifying uncertainty regarding how close it is to the decision boundary. For a given model, MT first obtains a set of mutated models through mutation operators, such as Gaussian Fuzzing (GF), Weight Shuffling (WS), then calibrates predictive distribution as the average logit over all mutated models. The UE method is defined as the label change rate (LCR).
    \item \textbf{Dissector (DS)} \citet{wang2020dissector} train a classification layer (snapshot) $S_l$ after each intermediate layer $l$ using linear regression, and estimates uncertainty by assuming a correctly classified input should induce an increasing confidence across the hidden layers. We compute the average logits over the resultant snapshots as calibrated outputs. The uncertainty is quantified by snapshot-profile-validity (SPV), whose weight parameters are formulated in three growth types: linear, logarithmic (log), and exponential (exp).
\end{itemize}

\begin{table*}[]
    \centering
    \resizebox{0.97\textwidth}{!}{
    \begin{tabular}{l|l||cccc|cc||cccc|cc}
    \toprule
        \multirow{2}{*}{\textbf{Shift Type}} & \multirow{2}{*}{\textbf{Method}} & \multicolumn{6}{c||}{\textbf{Code Completion}} & \multicolumn{6}{c}{\textbf{Code Summarization}} \\
        \cline{3-8}\cline{9-14}
        & & \textbf{Dev} & \textbf{Shift1} & \textbf{Shift2} & \textbf{Shift3} & \textbf{Avg}$\uparrow$ & \textbf{Avg Rank}$\downarrow$ & \textbf{Dev} & \textbf{Shift1} & \textbf{Shift2} & \textbf{Shift3} & \textbf{Avg}$\uparrow$ & \textbf{Avg Rank}$\downarrow$ \\
        \hline\hline
        \multirow{5}{*}{\textbf{\textsc{Timeline}}} & Base / TS & 71.87 & 71.84 & 69.25 & 67.46 & 70.11 & 4.75 & 48.04 & 47.82 & 46.98 & 46.89 & 47.43 & 5.00 \\
        & MCD & 73.54 & 72.68 & 71.38 & 69.40 & \textbf{71.75} & \textbf{1.25} & 48.83 & 48.57 & 47.87 & 47.38 & \underline{48.16} & \underline{2.50} \\
        & DE & 72.60 & 71.74 & 71.30 & 70.64 & \underline{71.57} & \underline{2.25} & 49.58 & 49.54 & 48.66 & 48.63 & \textbf{49.10} & \textbf{1.00} \\
        & MT & 72.02 & 71.84 & 69.28 & 68.47 & 70.40 & 3.5 & 48.20 & 47.95 & 47.14 & 47.24 & 47.63 & 4.00 \\
        & DS & 72.92 & 71.31 & 69.50 & 67.51 & 70.31 & 3.25 & 48.30 & 48.42 & 48.27 & 47.51 & 48.13 & \underline{2.50} \\
    \midrule
        \multirow{5}{*}{\textbf{\textsc{Project}}} & Base / TS & 68.05 & 65.76 & 65.46 & 64.76 & 66.01 & 4.75 & 53.64 & 52.86 & 48.63 & 46.84 & 50.49 & 4.50 \\
        & MCD & 69.79 & 65.81 & 65.60 & 64.82 & \underline{66.51} & \underline{2.50} & 54.53 & 54.13 & 49.12 & 47.24 & \underline{51.26} & \underline{1.75} \\
        & DE & 70.18 & 67.79 & 66.71 & 65.70 & \textbf{67.60} & \textbf{1.00} & 54.46 & 53.44 & 49.91 & 48.80 & \textbf{51.65} & \textbf{1.50} \\
        & MT & 68.14 & 65.77 & 65.54 & 64.96 & 66.10 & 3.00 & 54.12 & 53.02 & 48.65 & 47.99 & 50.95 & 3.00 \\
        & DS & 68.98 & 65.79 & 65.36 & 64.70 & 66.21 & 3.75 & 53.46 & 52.79 & 48.89 & 47.02 & 50.54 & 4.25 \\
    \midrule
        \multirow{5}{*}{\textbf{\textsc{Author}}} & Base / TS & 73.69 & 72.97 & 72.77 & 71.43 & 72.72 & 4.75 & 50.94 & 49.32 & 48.13 & 44.50 & 48.22 & 4.00 \\
        & MCD & 74.27 & 73.95 & 72.84 & 71.52 & \underline{73.15} & \underline{2.00} & 51.87 & 50.29 & 48.30 & 46.82 & \textbf{49.32} & \textbf{2.25} \\
        & DE & 75.07 & 74.71 & 73.23 & 72.55 & \textbf{73.89} & \textbf{1.00} & 51.17 & 49.99 & 48.74 & 46.23 & \underline{49.03} & \textbf{2.25} \\
        & MT & 73.75 & 73.11 & 73.08 & 71.49 & 72.86 & 3.25 & 50.95 & 49.37 & 48.14 & 45.54 & 48.50 & 3.50 \\
        & DS & 73.63 & 72.99 & 72.64 & 71.55 & 72.70 & 4.00 & 51.02 & 49.43 & 48.24 & 46.07 & 48.69 & \underline{3.00} \\
    \bottomrule
    \end{tabular}}
    \caption{F-1 (\%) and ranking of different methods for CodeLlama across all datasets. 
    All methods consistently produce lower F-1 under intensifying shifts.
    Probabilistic methods (except TS) outperform the vanilla baseline.}
    \label{tab:accuracy}
    \vspace{-0.3cm}
\end{table*}

To evaluate the performance of these methods, we consider two most-studied code analysis tasks (more details about task preprocessing and evaluation are described in Appendix~\ref{sec:task_process_eval}):
\begin{itemize}[leftmargin=*]
    \item \textbf{Code Completion~(CC)} We consider token-level CC~\cite{codetree}, which is analogous to language modeling of code generation. Specifically, we randomly mask a token in each code snippet as the missing part and feed the preceding context to CodeLlama. We evaluate the top-$k$ sub-token F-1 of the predicted masked tokens.
    \vspace{-0.3cm}
    \item \textbf{Code Summarization~(CS)} We focus on method name prediction~\cite{code2seq,jain-etal-2021-contrastive} which aims to describe names for code bodies (functions, classes, etc.). It is a challenging yet vital part of readable and maintainable code~\cite{nie-etal-2022-impact}. We evaluate the top-$k$ sub-token F-1 of the predicted names.
\end{itemize}




\subsection{Uncertainty Calibration Quality}

\textbf{Prediction Performance}
We first compare the post-calibration prediction accuracy of different methods.
We follow standard training, validation and testing (dev) protocols, but further evaluate results on increasingly shifted data (shift1 $\rightarrow$ shift3).
Table~\ref{tab:accuracy} summarizes the sub-token F-1 scores in two tasks, we observe that, despite the degradation of calibration quality due to intensifying shifts, probabilistic methods consistently mitigate such impact, as evidenced by consistently improved F-1 compared to the deterministic baseline (and TS) across shifted datasets.
Among them, DE and MCD achieve higher overall F-1 scores, while post-hoc calibration (TS) is more pone to degradation under distribution shifts. 
Other probabilistic methods also exhibit promising performance: 
MT, whose combined predictions can be viewed as another ensemble technique, consistently improves over the baseline and has relatively smaller F-1 drop in the shifted dataset.
Interestingly, \textsc{author shift} has less impact on CC quality compared to CS.
This suggests different programmers' design and implementation logic plays little role when decision making largely relies on \emph{code syntax}, such as common programming language grammars and rules in assisting CC.

\begin{figure*}
\centering
\includegraphics[width=0.95\textwidth]{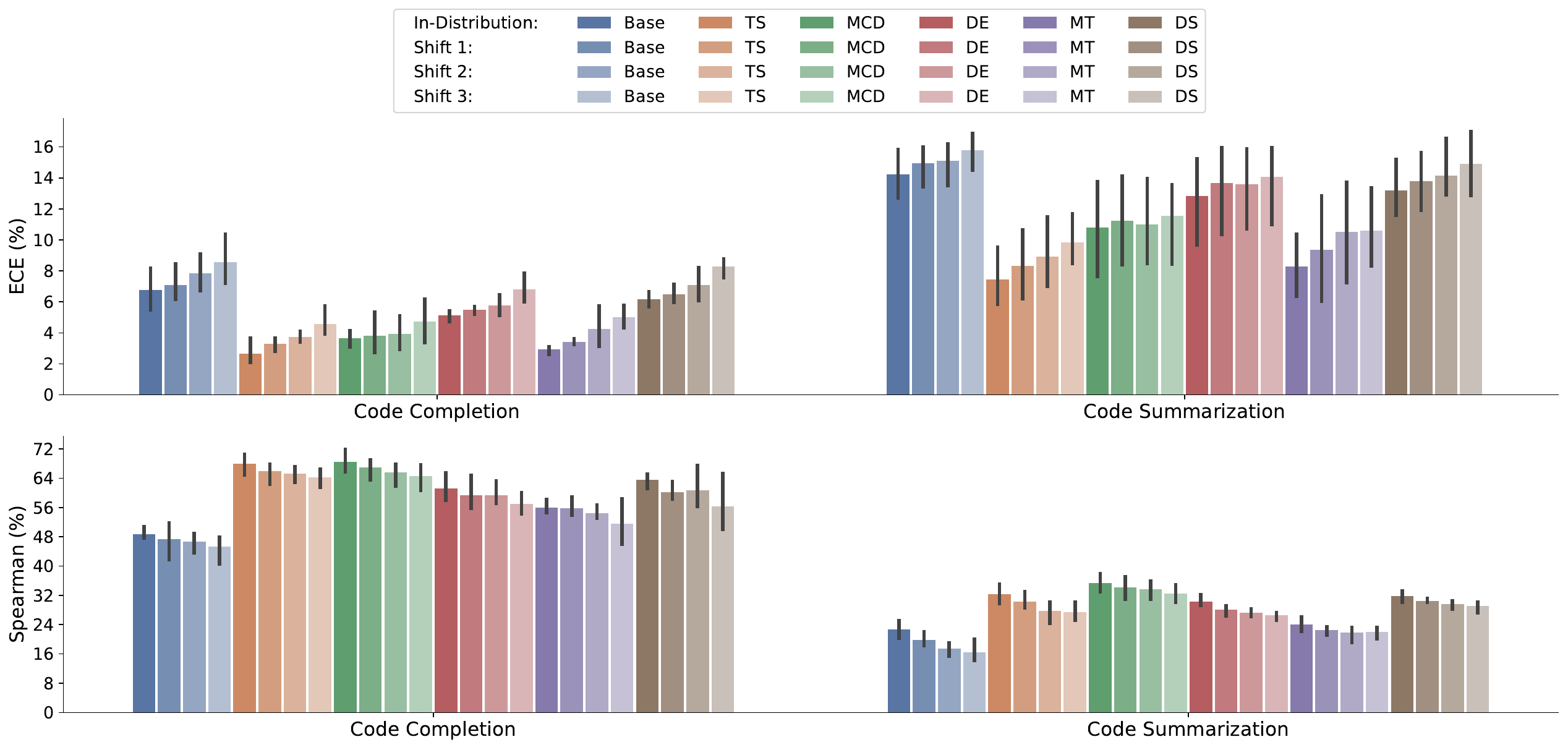}
\caption{ECE (\%) and Spearman's Rank Correlation (\%) for different uncertainty methods in both in-distribution and shifted code examples. Error bars indicate variations across the three shift patterns.}
\label{fig:calibration}
\vspace{-0.3cm}
\end{figure*}

\textbf{Expected Calibration Error (ECE)} 
We measure the difference in expectation between confidence and accuracy using ECE metric~\cite{ece} to evaluate how well the estimated model probabilities have been calibrated: 
\begin{equation}
    \text{ECE} = \sum_{k=1}^K\frac{|B_k|}{n}|\text{conf}(B_k) - \text{acc}(B_k)|
\end{equation}

Here we group $n$ sample predictions into $K$ interval bins and define $B_k$ as the set of indices of examples whose prediction confidence lie in the $k^{th}$ bin $B_k = (\frac{k-1}{K}, \frac{k}{K}]$, where its accuracy and confidence are defined as $\text{acc}(B_k) = \frac{1}{|B_k|}\sum_{i\in B_k}\mathbf{1}(\hat{y}_i = y_i)$ and $\text{conf}(B_k) = \frac{1}{|B_k|}\sum_{i\in B_k}\hat{p}_i$. 
In code analysis, $\hat{y}_i$ is the predicted token, and $\hat{p}_i$ is the corresponding probability.
As shown in Figure~\ref{fig:calibration}, probabilistic methods consistently outperform the deterministic baseline.
Among them, TS, MCD and MT better diminish the ECE, suggesting they facilitate to maintain the competitive prediction quality of CodeLlama under distribution shifts.
DE exhibits slightly improved ECE quality over the baseline, while DS falls short in reducing calibration errors.


\textbf{Rank Correlation with Quality Score}
We further evaluate how the Spearman's rank correlation between the calibrated probabilities and accuracy changes with intensifying shifts.
As shown in Figure~\ref{fig:calibration}, we observe a general trend demonstrating the calibration of probabilistic methods increase the correlation under distribution shifts.
MCD and TS exhibit superior calibration capacity with the higher correlation scores, whereas MT demonstrates limitations in improving correlation over baseline.

\subsection{Uncertainty Estimation~(UE) Precision}

To comprehensively assess the quality of UE methods, We consider three widely-applied metrics:
\begin{itemize}[leftmargin=*]
    \item \textbf{Area Under the ROC curve (AUC)}$\uparrow$ \citet{hendrycks2016baseline} suggest evaluating the quality of UE using AUC, which can be interpreted as the probability that a misclassified example has a greater uncertainty score than a correctly-classified one. 
    \vspace{-0.3cm}
    \item \textbf{Area Under the Precision-Recall curve (AUPR)}$\uparrow$ To better handle the situation when the positive class and negative class have greatly differing base rates, AUPR is also suggested to evaluate the quality of UE. The PR curve plots the relationship between precision and recall.
    \vspace{-0.3cm}
    \item \textbf{Brier}$\downarrow$ \citet{brier1950verification} measure the mean squared error of the uncertainty scores assigned to each sample and the actual outcome.
\end{itemize}

\begin{figure*}
    \centering
    \includegraphics[width=0.95\textwidth]{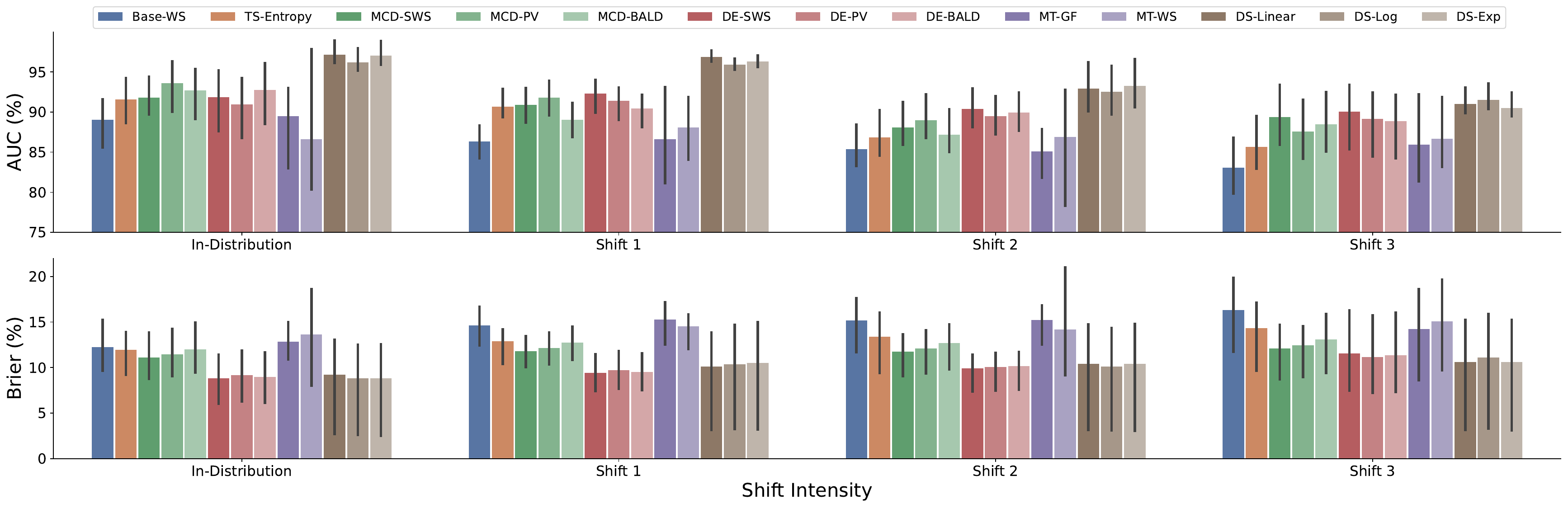}
    \caption{Misclassification detection results for probabilistic methods with corresponding UE techniques in CC under intensifying shifts. Error bars indicate variations across the three shift patterns. Results in CC and AUPR results in both tasks are demonstrated in Figure~\ref{fig:mis_detect_cs} and Figure~\ref{fig:mis_detect_aupr} in Appendix~\ref{sec:additional_results}.}
    \label{fig:mis_detect_cc}
\end{figure*}


\textbf{Misclassification Detection} 
In this experiment, we evaluate how the uncertainty scores ($u\in[0,1]$) of different UE techniques correlate with the mistakes $\bar{e} = \mathbf{1}(y_i \neq \hat{y}_i)$ of the CodeLlama under distribution shifts.
An effective UE method should produce higher scores for mistakes.
Figure~\ref{fig:mis_detect_cc} presents the misclassification detection performance of different UE methods across all three shift patterns.
We observe a general trend of decreasing UE quality as shift intensity grows, as evidenced by lower AUC and Brier scores for all methods in the shift3 set.
Probabilistic methods generally outperform the baseline, cautioning the significance of calibration.
Among them, DS achieves the highest AUC and rather low Brier scores under distribution shifts, suggesting its assumption of cross-layer consistency is effective in detecting the mistakes of LLMs.
Ensemble method (DE) not only shows promising results but also exhibit relative robustness to intensifying shifts.
Whereas adversarial method (MT) falls short in misclassification detection, producing even lower quality than the baseline in some cases. 
This indicates its adversarial assumption may not necessarily apply to mistake detection and cause false alarms.
Also, MT is more sensitive to different shift patterns, as evidenced by larger error bars in both AUC and Brier.

\begin{figure*}
\centering
\includegraphics[width=0.98\textwidth]{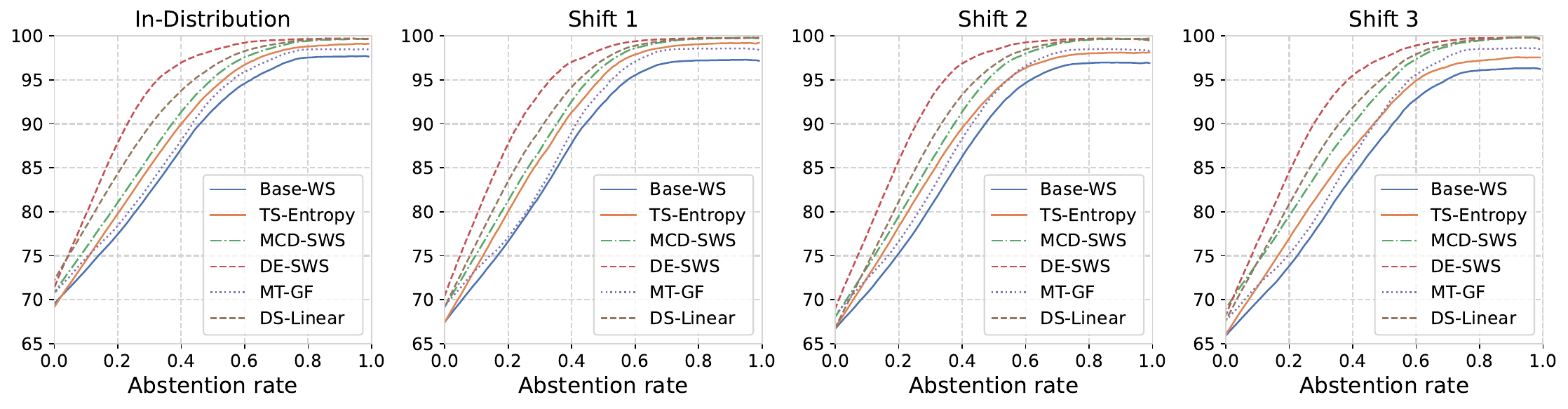}
\caption{F-1 (\%) vs. Abstention curve for different uncertainty methods in CC under distribution shifts. For each method, we report one corresponding UE results. Each line is the average F-1 score over the three shift patterns. Results in CS and remaining UE results are shown in Figure~\ref{fig:filter_cs} and Figure~\ref{fig:filter_remain} in Appendix~\ref{sec:additional_results}.}
\label{fig:filter_cc}
\vspace{-0.3cm}
\end{figure*}


\textbf{Selective Prediction} 
Selective prediction refers to selectively predicting high-quality outputs while rejecting the low-quality outputs~\cite{llm-ood-selection}. 
It evaluates how effective and efficient existing methods can improve the performance requirement by allowing a model to abstain from predicting highly-uncertain instances, assuming an effective UE method indicates high uncertainty for low-quality outputs. 
As shown in Figure~\ref{fig:filter_cc}, at a given abstention rate $\tau$, we filter out the lowest $\tau$-fraction uncertain examples based on the estimated scores and compute the average F-1 score of the remaining examples.
We observe all probabilistic methods improve the quality of CodeLlama to large extent, especially in the shift3 set.
Similar to misclassification detection, DE and DS are more efficient: they achieve higher F-1 by rejecting the same proportion of code examples. 
Whereas adversarial method slightly improves the baseline.

\textbf{OOD Detection}
Besides the three code distribution shifts, we further evaluate the UE quality under a more severe shift, projects that are designed for \emph{different programming paradigms} (\textsc{paradigm shift}).
We consider two real-world projects, BigCloneEval\footnote{\url{https://github.com/jeffsvajlenko/BigCloneEval}} (in-distribution) and Ruoyiplus\footnote{\url{https://github.com/kongshanxuelin/ruoyiplus}} (OOD).
The former is a Java language tool for analyzing code clone detection, while the latter is a J2EE development system for backend management.  
We extract Java files from both projects as raw snippets.
The preprocessed dataset statistics is detailed in Appendix~\ref{sec:ood_data}.
Table~\ref{tab:ood} summarizes the calibrated prediction accuracy and OOD detection quality of CodeLlama for different methods.
We observe an improved F-1 score of probabilistic methods, although the overall quality is much lower in the OOD dataset compared to the three shifted datasets.
Among them, DE and MCD still show promising calibration effectiveness.
Interestingly, adversarial method (MT) consistently outperform others in detecting OOD code examples, with higher AUC, AUPR and lower Brier scores.
This suggests its adversarial assumption effectively applies to severely shifted inputs when considering their distance to the model's decision boundary.
DS also achieves top-rank results in OOD detection, which proves the sensitivity of its UE to distribution shifts.

\begin{table*}[]
    \centering
    \resizebox{0.87\textwidth}{!}{
    \begin{tabular}{l|l||c|cccc||c|cccc}
    \toprule    
       \multirow{2}{*}{\textbf{Method}} & \multirow{2}{*}{\textbf{UE}} & \multicolumn{5}{c||}{\textbf{Code Completion}} & \multicolumn{5}{c}{\textbf{Code Summarization}} \\
       \cline{3-7}\cline{8-12}
       & & \textbf{F-1}$\uparrow$ & \textbf{AUC}$\uparrow$ & \textbf{AUPR}$\uparrow$ & \textbf{Brier}$\downarrow$ & \textbf{Avg Rank}$\downarrow$ & \textbf{F-1}$\uparrow$ & \textbf{AUC}$\uparrow$ & \textbf{AUPR}$\uparrow$ & \textbf{Brier}$\downarrow$ & \textbf{Avg Rank}$\downarrow$ \\
    \midrule
    \midrule
        Base &  WS & 55.48 & 69.99 & 78.18 & 23.78 & 8.67 & 29.99 & 59.63 & 69.47 & 22.67 & 10.67 \\
        TS & Entropy & 55.48 & 70.68 & 78.56 & 25.31 & 9.00 & 29.99 & 60.18 & 70.30 & 25.84 & 11.67 \\
        \midrule
        \multirow{3}{*}{MCD} & SWS & \multirow{3}{*}{\underline{57.18}} &  69.25 & 81.71 & 20.34 & 4.67 & \multirow{3}{*}{\underline{33.32}} & 64.84 & 72.80 & 24.27 & 9.00 \\
        &  PV & & 73.12 & 77.08 & 31.96 & 10.67 &  & 65.06 & 74.25 & 22.13 & 6.67  \\
        &  BALD & & 72.62 & 77.34 & 30.76 & 10.33 &  & 67.79 & 71.81 & 24.03 & 7.67 \\
        \midrule
        \multirow{3}{*}{DE} & SWS & \multirow{3}{*}{\textbf{58.79}} & 66.53 & 79.94 & 24.34 & 8.33 & \multirow{3}{*}{\textbf{35.15}} & 63.97 & 78.94 & 25.78 & 9.00 \\
        & PV &  & 63.78 & 79.41 & 25.83 & 10.67 &  & 62.45 & 80.81 & 28.01 & 9.00  \\
        & BALD &  & 64.29 & 79.55 & 25.36 & 9.67 &  & 61.92 & 79.30 & 26.33 & 9.67 \\
        \midrule
        \multirow{2}{*}{MT} & GF  & \multirow{2}{*}{56.86} & 77.24 & 80.69 & 19.28 & \underline{2.67} & \multirow{2}{*}{30.12} & 75.38 & 83.42 & 17.81 & \textbf{1.33} \\
        & WS &  & 78.25 & 81.33 & 17.50 & \textbf{1.33} & & 73.80 & 85.65  & 19.49 & \underline{1.67} \\
        \midrule
        \multirow{3}{*}{DS} & SPV-Linear & \multirow{3}{*}{57.04} & 73.63 & 79.41 & 20.44 & 6.00 & \multirow{3}{*}{30.19} & 68.62 & 79.27 & 21.35 & 4.33 \\
        & SPV-Log &  & 73.79 & 80.83 & 23.96 & 4.67 &  & 66.73 & 80.39 & 22.73 & 6.00  \\
        & SPV-Exp &  & 74.89 & 80.26 & 21.37 & 4.33 &  & 67.46 & 81.08 & 22.16 & 4.33  \\
    \bottomrule
    \end{tabular}}
    \caption{Prediction F-1 (\%) of CodeLlama on OOD examples and OOD detection performance (AUC, AUPR, Brier, Avg Rank) for different methods.}
    \label{tab:ood}
    \vspace{-0.3cm}
\end{table*}


\section{Further Analysis}

\subsection{Empirical Findings}

\begin{table}[]
    \centering
    \resizebox{0.45\textwidth}{!}{
    \begin{tabular}{l|ccc|ccc|c}
    \toprule
        \multirow{2}{*}{\textbf{Method}} & \multicolumn{3}{c|}{\textbf{Calibration}} & \multicolumn{3}{c|}{\textbf{UE}} & \multirow{2}{*}{\textbf{Score}$\uparrow$} \\
        \cmidrule{2-7}
        & \textbf{F-1} & \textbf{ECE} & \textbf{Corr} & \textbf{MD} & \textbf{SP} & \textbf{OOD} & \\
    \midrule
        Base & ✗ & ✗ & ✗ & ✗ & ✗ & ✗ & \emph{0.0} \\
        TS & ✗ & ✓✓ & ✓✓ & ✓ & ✓ & ✗ &  6 \\
        MCD & ✓✓ & ✓✓ & ✓✓ & ✓ & ✓ & ✗ & \textbf{8} \\
        DE & ✓✓ & ✓ & ✓ & ✓✓ & ✓✓ & ✗ & \textbf{8} \\
        MT & ✓ & ✓✓ & ✗ & ✗ & ✓ & ✓✓ & 6 \\
        DS & ✓ & ✗ & ✓ & ✓✓ & ✓✓ & ✓ & \underline{7} \\
    \bottomrule
    \end{tabular}}
    \caption{Overall comparison of studied methods. MD, SP are abbreviations of misclassification detection and selective prediction. ✓✓ denotes top-ranked, ✓ presents above-baseline, and ✗ means equal/below-baseline.}
    \label{tab:empirical_findings}
\end{table}

Table~\ref{tab:empirical_findings} summarizes the overall performance of studied methods in terms of both calibration ability and UE quality. 
For calibration ability, we observe MCD achieves the overall best performance with top-ranked F-1 and calibration quality.
Other probabilistic methods also show promising results: DE is better at prediction F-1, while MT and TS exhibit better calibration quality.
DS, in contrast, falls short in calibrating CodeLlama under distribution shifts.
For UE quality, however, we see DS produces the overall best results, with top-ranked low-quality output detection precision (i.e., detecting mistakes and abstain) and distribution sensitivity (i.e., detecting OOD samples).
Other probabilistic methods also improves UE quality compared to the baseline:
DE is superior in misclassification detection and selective prediction, while MT is more sensitive to distribution shifts.
These performance gap between different criteria suggests the hypotheses of certain methods fit in specific scenarios with loss of generalizability.

\subsection{Efficiency Analysis}

\begin{table}[]
    \centering
    \resizebox{0.49\textwidth}{!}{
    \begin{tabular}{lcccc}
    \toprule    
       \textbf{Method}  & \textbf{Calibration}$\downarrow$ & \textbf{UE}$\downarrow$ & \textbf{Total}$\downarrow$ & \textbf{Snippet}$\downarrow$\\
    \midrule
        Base & \emph{0.0} & 65 & 65 ($\times$1) & \textbf{0.27}   \\
        TS & 69 & 75 & 144 ($\times$2) & \underline{0.61} \\
        MCD* & 5973 & 157.83* & 6130.83 ($\times$94) & 25.85 \\
        DE* & 3162 & 87.39* & 3249.39 ($\times$50) & 13.50 \\
        MT* & 7917 & 132.26* & 8049.94 ($\times$124) & 33.94  \\
        DS* & 135 & 71.39* & 206.39 ($\times$3) & 0.87 \\
        
    \bottomrule
    \end{tabular}}
    \caption{Overhead (s) for different probabilistic methods. Per snippet ($\times 10^{-3}$s) is the instance-level overhead. For method* containing multiple UE results, we report the average number as they cost similar overhead.}
    \label{tab:overhead}
    \vspace{-0.3cm}
\end{table}

Table~\ref{tab:overhead} compares the overhead of different methods, where the calibration overhead of DE is the averaged per epoch overhead for training $T_{\text{de}}$ independent CodeLlama models.
The hardware setting is detailed in Appendix~\ref{sec:hardware}.
We observe a general trade-off between efficacy and efficiency: methods requiring sophisticated calibration such as MCD, DE, generally produce promising results but cost more overhead.
Their practical overhead may even inflate: the complexity of DS is proportional to the number of hidden layers used for training snapshots.
In time-sensitive tasks such as online code generation, TS or vanilla baseline may be preferred for selective prediction, especially when deploying computation-intensive models.

\section{Conclusion}

LLMs are prone to performance degradation under various code distribution shifts.
This study identifies three real-world code distribution shift patterns that adversely affect CodeLlama's prediction quality.
We investigate five cutting-edge probabilistic methods, focusing on both their calibration abilities and uncertainty estimation~(UE) efficacy in these shifting contexts.
Our findings reveal that these methods generally alleviate the negative effects of distribution variations, leading to improved prediction accuracy and higher precision of low-quality output detection.
However, our analyses also uncover performance variation across different evaluation criteria and various shift intensities, due to limited application scenarios of method hypotheses. 
For instance, adversarial assumption (MT) aligns well with severely OOD instances but little with misclassified ones; ensemble method is more robustness against severe shifts compared to post-hoc calibration.
We further identify a general trade-off between efficacy and efficiency, cautioning the importance of choosing appropriate method under specific circumstances.

\section*{Limitations}
This study investigates the uncertainty awareness of LLMs under various code distribution shifts, using advanced probabilistic methods.
While our benchmark datasets mainly focus on the Java programming language, the scope of real-world code distribution shifts covers a variety of other languages, such as Python and C++.
Our methodologies for dataset creation and experimental designs hold the potential for broad applicability across these diverse programming contexts.
Furthermore, our current research regards code analysis tasks as token-level classification (although token-level CC is analogous to language modeling). 
In our future work, we aim to extend our evaluation to generation tasks, such as comment generation and code search, to lead to more comprehensive understanding of LLM's capabilities. 
However, these generation tasks typically require human annotations when creating shifted datasets (e.g., comments that include time-sensitive information may need human calibrations). 
This requirement underscores the need for more automated approaches to address distribution shifts in the future.

\section*{Acknowledgements}
This research was supported by the National Science Foundation under Grants CNS Career 2230968, CPS 2230969, CNS 2300525, CNS 2343653, CNS 2312397.


\bibliography{anthology,custom}

\appendix



\section{Settings}

\subsection{Preprocessing and Evaluation}

\label{sec:task_process_eval}

\textbf{Preprocessing}
Each Java file is converted into a string for tokenization and preprocessing. 
We normalize uncommon literals for better user experience. 
As suggested by ~\cite{codexglue}, we don't require models to identify literals such as names, IP address, phone numbers or numeric literals, and thus normalize them by pre-defined special tokens. 
Considering that frequently-used literals may contain useful information, e.g., ``\_\_main\_\_'' or ``utf-8'', we preserve the 200 most frequent string and 30 most frequent numeric literals. 
These literals will be normalized by tokens in ``<STR\_LIT:utf-8>'' format, while uncommon literals are replaced by <STR\_LIT> or <NUM\_LIT>.
We add <s> and </s> to indicate the start and the end of one snippet of code, and <MASK> one token for prediction in each snippet line.

\textbf{Code Completion} 
CC aims to predict next code token given context of previous tokens, is a one of the most widely used features in software development through IDEs. 
An effective code completion tool could improve software developers' productivity.
We evaluate CodeLlama in terms of sequence classification by top-$k$ sub-token F-1 score following~\cite{codexglue}, which is analogous to language modeling. 
For instance, consider the function ``count|lines'' with $k=2$. 
We tokenize it into sub-tokens: ``count'' and ``lines'', using a Llama pre-trained tokenizer.
Then we query WordNet~\cite{miller1995wordnet} to obtain the top-2 synonyms for each sub-token: ``tally'', ``total'' (for ``count''), and ``rows'', ``stripes'' (for ``lines'').
Combinations of synonyms like ``lines|tally'' and ``count|rows'' are treated as exact matches.
Partial matches like ``count'' implies full precision but low recall, while extra matches like \emph{CountBlankLines} suggests full recall but low precision. 
Out-of-vocabulary~(OOV) sub-tokens are treated as false negatives, which reduces the recall performance.

\textbf{Code Summarization} Existing studies regarding CS consists of two main tasks: comment generation and method name prediction. In this work, we focus on MNP, which aims to predict method names based on the context provided by code snippets. Similar to CC, function names are also composed of sub-tokens which implies non-unique ground truths, and we evaluate the top-2 sub-token F-1 score of the predicted method names.

\subsection{Model Setup}
\label{sec:model_implementation}
We use the Hugging Face pre-trained model \emph{codellama/CodeLlama-7b-hf}.
For each shifted dataset, the hyperparameters of CodeLlama is tuned using grid search.
We train
\footnote{Our code is available at \url{https://github.com/yul091/LLMUncertainty}.} 
CodeLlama for 100 epochs, with a learning rate of 2e-5 and weight decay of 1e-4, using the AdamW optimizer~\cite{adamW}.
The hidden dimension $d_h$ is set as 256, and the training batch size is 8.
For each experiment, we run the model 5 times and report the average numbers as the evaluation results.
Below are pre-defined list from which we select hyperparameter values:

\noindent\textbf{Learning rate}: [5e-6, 7e-9e-6, 1e-5, 2e-5, 5e-5, 7e-5, 1e-4, 2e-4, 5e-4];

\noindent\textbf{Number of epochs}: $\left \{ n\in\mathbb{N}| 15 \leq n \leq 150 \right \}$;

\noindent\textbf{Batch size}: [2, 4, 6, 8, 10, 12, 14, 16];

\noindent\textbf{Weight decay}: [0, 1e-5, 1e-4, 1e-3, 1e-2, 1e-1];

\noindent\textbf{Hidden dim.}: [32, 64, 128, 256, 512, 1024].

\subsection{OOD Datasets}

\label{sec:ood_data}

\begin{table}[]
    \centering
    \resizebox{0.4\textwidth}{!}{
    \begin{tabular}{lcc}
    \toprule
     & \textbf{BigCloneEval} & \textbf{Ruoyiplus} \\
     \midrule
    Paradigm & Language tool & J2EE platform \\
    Snippet size & 10.74/10.73 & 9.01 \\
    \# Snippets & 15,652/3,896 & 3,768 \\
    KL$\uparrow$ & \emph{0.0} & 3.44 \\
    Cosine$\downarrow$ & \emph{0.0} & 0.68 \\
    Vocab & 13,828 & 11,704 \\
    \bottomrule
    \end{tabular}}
    \caption{Statistics of the BigCloneEval (in-distribution) and Ruoyiplus (OOD) datasets.}
    \label{tab:ood_dataset}
\end{table}

Both BigCloneEval and Ruoyiplus are relatively large datasets, and we extract 15,000 Java files in both projects.
We random select 2,000 / 500 Java files from BigCloneEval as the training / dev datasets, and 500 Java files from Ruoyiplus as the OOD testset.
Table~\ref{tab:ood_dataset} shows their statistics after preprocessing, where we see a much severer shift compared to the previous three according to the higher KL and lower cosine scores.

\subsection{Hardware}

\label{sec:hardware}

\begin{table}[]
    \centering
    \resizebox{0.49\textwidth}{!}{
    \begin{tabular}{lc}
    \toprule
        CPU & Intel(R) Xeon(R) Silver 4314 CPU @ 2.40GHz \\
        CPU Cores & 32 \\
        GPU & NVIDIA RTX 6000 Ada Generation \\
        GPU Memory & 50 GB \\
    \bottomrule
    \end{tabular}}
    \caption{Hardware configuration in the experiments.}
    \label{tab:hardware}
\end{table}

All of our experiments, including the fine-tuning of CodeLlama, uncertainty calibration, and evaluation, are conducted on an NVIDIA A6000 Ada Server, as shown in Table~\ref{tab:hardware}.

\section{Probabilistic Methods}

\label{sec:uncertainty_setup}

\begin{table*}[]
    \centering
    \resizebox{0.85\textwidth}{!}{
    \begin{tabular}{lll}
    \toprule
        \textbf{Mutation Operator} & \textbf{Level} & \textbf{Description}  \\
         \midrule
        Gaussian Fuzzing (GF) & Weight & Fuzz weights of CodeLlama by Gaussian Distribution \\
        Weight Shuffling (WS) & Neuron & Shuffle selected weights with ratio 25\% \\
        Neuron Switch (NS) & Neuron & Switch two neurons within each layer of CodeLlama \\
        Neuron Activation Inverse (NAI) & Neuron & Deactivate the status of a neuron with ratio 25\% \\
    \bottomrule
    \end{tabular}}
    \caption{Model mutation operators in Mutation Testing.}
    \label{tab:mutation_operators}
\end{table*}

\begin{table*}[]
    \centering
    \resizebox{0.8\textwidth}{!}{
    \begin{tabular}{lllc}
    \toprule
        \textbf{Grow Type} & \textbf{Formula} & \textbf{Reduced Formula} & \textbf{Number of Parameters}  \\
         \midrule
        Linear & $\alpha = ax+l$ & $\left\{
            \begin{array}{ll}
            \alpha = x \\
            \alpha = wx+1
        \end{array}
        \right.$ & 2 $\rightarrow$ 1 \\
        Logarithmic & $\alpha = a\log_b(lx+k_1) + k_2$ & $\left\{
            \begin{array}{lll}
            \alpha = \ln x \\
            \alpha = w\ln x+1 \\
            \alpha = w\ln(\beta x + 1) + 1
        \end{array}
        \right.$ & 5 $\rightarrow$ 2 \\
        Exponential & $\alpha = ae^{lx+b_1}+b_2$ & $\left\{
            \begin{array}{ll}
            \alpha = e^{\beta x} \\
            \alpha = we^{\beta x}+1
        \end{array}
        \right.$ & 4 $\rightarrow$ 2 \\
    \bottomrule
    \end{tabular}}
    \caption{Parameter reduction for modeling weights in Dissector.}
    \label{tab:pv_formula}
\end{table*}

In this section, we illustrate the additional details for the probabilistic methods, providing a comprehensive understanding of their individual characteristics and how their properties may facilitate in LLM calibration.
For Monte-Carlo Dropout, Deep Ensemble, Mutation Testing, and Dissector, the essential idea is to get an approximation of the predictive posterior of a model through averaging the probabilities from several independent models (also referred to the \emph{committee size}).

\textbf{Vanilla Baseline} Vanilla logits (after softmax) are directly used as predictive probabilities without calibration.
The UE method is winning score:

\begin{equation}
    u_{\text{ws}}(x) = 1 - \max_{c\in C}p(y=c|x)=1-\max_{c\in C}p_t^c
\end{equation}

The time complexity is $O(B)$, where $B$ is the complexity of base model, e.g., CodeLlama.

\textbf{Temperature Scaling} TS is a post-hoc calibration method based on a hold-out validation set.
It works effectively on in-distribution sets, but can be prone to degradation under distribution shifts, as shown by our study.
We tune the parameter $T_{\text{ts}}$ using a validation set, by minimizing the negative log-likelihood (NLL) loss between the scaled logits and labels with the LBFGS optimizer for 50 training epochs. The time complexity is $O(1) + O(B)=O(B)$.
The UE method is Shannon entropy over the new predictive distribution: 

\begin{equation}
    u_{\text{entropy}} = -\sum_{c=1}^{C} \frac{e^{l^c/T_{\text{ts}}}}{\sum_k{e^{l^k/T_{\text{ts}}}}} \log\left(\frac{e^{l^c/T_{\text{ts}}}}{\sum_k{e^{l^k/T_{\text{ts}}}}}\right)
\end{equation}

\textbf{Monte-Carlo Dropout} MCD is a Bayesian technique leveraging dropout regularization during training inference. 
By sampling multiple predictions, MCD measures how \emph{inner-model} parameters align with the data distribution and thus the confidence in its predictions.
We set the number of dropout samples $T_{\text{mcd}}$ as 10 and the dropout rate as 0.1.
The time complexity is $O(B \times T_{\text{mcd}})$.
The three UE methods are formulated as:

\begin{equation}
    u_{\text{sws}} = 1 - \frac{1}{T_{\text{mcd}}}\sum_{t=1}^{T_{\text{mcd}}}\max_{c\in C}p_t^c
\end{equation}

\begin{equation}
    u_{\text{pv}} = \frac{1}{C}\sum_{c=1}^C\left ( \frac{1}{T_{\text{mcd}}}\sum_{t=1}^{T_{\text{mcd}}}(p_t^c - \overline{p_t^c})^2 \right )
\end{equation}

\begin{equation}
    u_{\text{bald}} = \frac{1}{T_{\text{mcd}}}\sum_{t,c}p_t^c\log p_t^c - \frac{1}{C}\sum_{c=1}^C\overline{p_t^c}\log \overline{p_t^c}
\end{equation}

\textbf{Deep Ensemble}. 
DE is a non-Bayesian method that trains several independent deterministic models, with each capture different features of the data and output more precise and robust predictions when combined.
Its uncertainty captures the \emph{cross-model} parametric uncertainty.
Our study also proves its robustness in both top-ranked calibration ability and uncertainty estimation precision.
We train $T_{\text{de}} = 5$ deterministic CodeLlama models and averages all. In the original work, each model is trained with different architectures or subsets or training data.
We focus on CodeLlama and apply self-ensemble~\cite{wang-etal-2022iclr-ensemble} in our experiments, which has been shown to be helpful for boosting LLM performance~\cite{li-etal-2023-uncertainty}.
The time complexity is $O(B \times T_{\text{de}})$.
The three UE methods are the same as MCD but across ensemble models.

\textbf{Mutation Testing} 
MT assumes adversarial inputs are closer to the model decision boundary and thus more sensitive to the model mutation operations. 
\cite{wang2019adversarial} propose four DNN model mutation operators, GF, WS, NS, NAI, as shown in Table~\ref{tab:mutation_operators}.
For each operator, we mutate 100 times to obtain a series of mutated models $\mathcal{M} = \left \{ f'_i | 1 \leq i \leq 100 \right \}$ where $|\mathcal{M}|=100$. 
In other words, the calibrated predictive distribution and UE results are obtained from the 100 mutated models.
To save space, we only report the UE results of MT-GF and MT-WS in the paper, as the other two techniques produce similar patterns.
The time complexity is $O(B + |\mathcal{M}| \times F)$, where $F$ is the complexity of each mutation operation.
As $|\mathcal{M}|$ is typically large and $F$ is related to the size of deep models, the calibration overhead of MT can be extremely large in practice.
The UE method is label change rate (LCR) between the original output and mutated model outputs:

\begin{equation}
    u_{\text{lcr}}(x) = \frac{|\left \{ f'_i(x)\in \mathcal{M} | f'_i(x) \neq f(x) \right \}|}{|\mathcal{M}|}
\end{equation}

\begin{table*}[htp]
    \centering
    \resizebox{0.7\textwidth}{!}{
    \begin{tabular}{l|cccc|cccc}
    \toprule
        \multirow{2}{*}{\textbf{Shift Type}} & \multicolumn{4}{c|}{\textbf{Code Completion}} & \multicolumn{4}{c}{\textbf{Code Summarization}} \\
        & \textbf{Dev} & \textbf{Shift1} & \textbf{Shift2} & \textbf{Shift3} & \textbf{Dev} & \textbf{Shift1} & \textbf{Shift2} & \textbf{Shift3} \\
        \midrule
        \textsc{Timeline} & 0.44 & 0.55 & 0.46 & 0.43 & 0.03 & 0.06 & 0.06 & 0.09 \\
        \textsc{Project} & 0.44 & 0.21 & 0.69 & 0.35 & 0.08 & 0.08 & 0.02 & 0.08 \\
        \textsc{Author} & 0.57 & 0.94 & 0.82 & 1.15 & 0.21 & 0.19 & 0.16 & 0.15 \\
        \midrule
        \textbf{Shift Type} & \textbf{Dev} & \multicolumn{3}{c|}{\textbf{OOD}} & \textbf{Dev} & \multicolumn{3}{c}{\textbf{OOD}} \\
        \midrule
        \textsc{Paradigm} & 0.41 & \multicolumn{3}{c|}{\emph{0.0}}  & 0.08 &  \multicolumn{3}{c}{0.12}  \\
    \bottomrule
    \end{tabular}}
    \caption{Zero-shot F-1 performance ($\times 10^{-2}$) of CodeLlama across all the shifted and OOD datasets.}
    \label{tab:zero_shot_F1}
\end{table*}

\begin{figure*}
    \centering
    \includegraphics[width=0.95\textwidth]{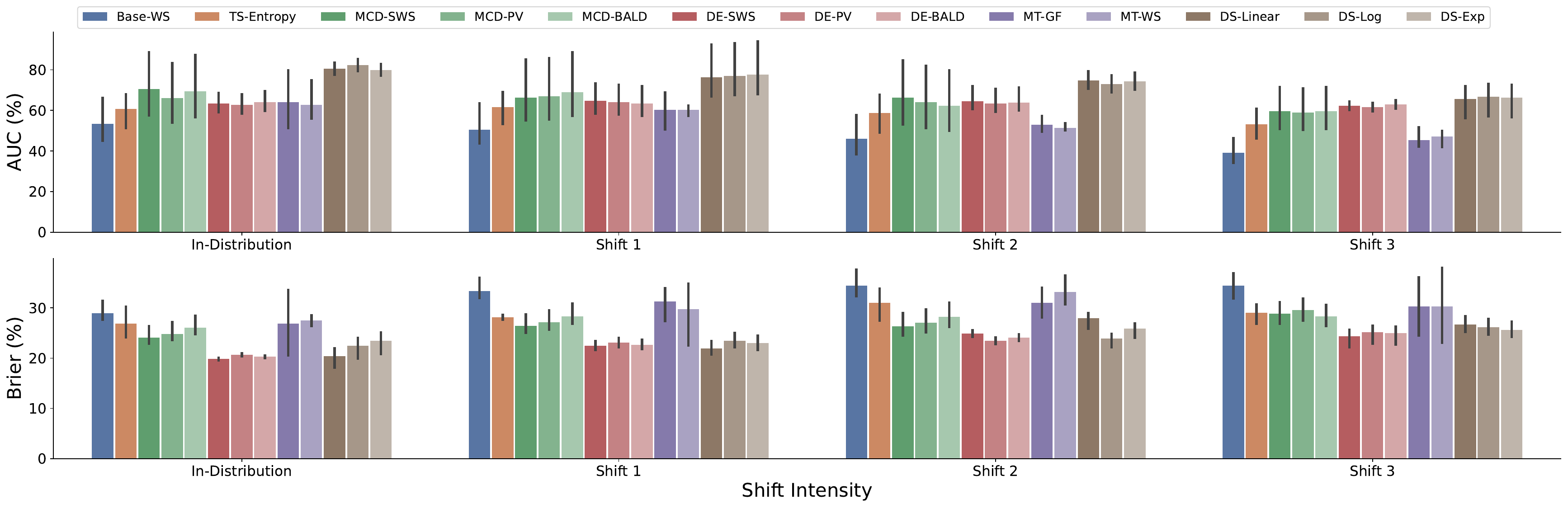}
    \caption{Misclassification detection results (AUC, Brier) for probabilistic methods with corresponding UE techniques in CS under intensifying shifts. Error bars indicate variations across the three shift patterns. }
    \label{fig:mis_detect_cs}
\end{figure*}

\textbf{Dissector} 
DS assumes a correctly-classified inputs should has an increasing intermediate consistent across hidden layers, and is thus effective in misclassification detection.
We select the 1st, 3rd, 5th, 7th, and 9th transformer layer of CodeLlama to train the corresponding snapshots.
Each snapshot is a linear layer trained using stochastic gradient descent~(SGD) with cross entropy loss for 10 epochs.
Suppose the original model $f$ predicts $x$ as label $\hat{y}$, the snapshot validity (SV) score first measures how $\hat{y}$ is uniquely supported by the probability vector in this snapshot.
If $\hat{y}$ is aligned with the highest probability, Dissector measures uniqueness as how much $\hat{y}$'s associated probability $p_{S_l}^{\hat{y}}$ exceeds the second highest probability $p_{S_l}^{c_{(2)}}$ (with label $c_{(2)}$).
Otherwise, Dissector measures how much the actual highest probability $p_{S_l}^{c_{(1)}}$ (with label $c_{(1)}$) exceeds that of $\hat{y}$:

\begin{equation}
    \text{SV}(\hat{y}, S_l) = \left\{
  \begin{array}{ll}
     \frac{p_{S_l}^{\hat{y}}}{p_{S_l}^{\hat{y}} + p_{S_l}^{c_{(2)}}},  & \hat{y} = \argmax\limits_{c} p_{S_l}^c \\
     1-\frac{p_{S_l}^{c_{(1)}}}{p_{S_l}^{\hat{y}} + p_{S_l}^{c_{(1)}}},  & \text{otherwise}
  \end{array}
  \right.
\end{equation}

The SV score is in range [0, 1], with a larger value indicating more uniquely the current snapshot supports the final prediction result $\hat{y}$.
The UE method is profile validity (PV), i.e., the normalized sum of SV scores across layers:

\begin{equation}
    u_{\text{spv}} = \frac{\sum_{l\in L} \alpha_l\cdot\text{SV}_l(\hat{y}, S_l)}{\sum_{l\in L}\alpha_l}
\end{equation}

Here $L$ is the set of hidden layers for training snapshots and $\alpha_l$ is associated weight for the $l^{th}$ layer. 
Table~\ref{tab:pv_formula} lists general formulas for computing weight $\alpha$ values regarding the three growth types, based on the DS assumption that a within-input should have an increasing confidence cross the intermediate layers of a model.
To reduce parameters, \cite{wang2020dissector} also reduce these formulas with one ($w$) or two ($w$ and $\beta$) parameters only.
The time complexity is $O(B + |L|\times W)$, where $L$ is the set of hidden layers used for training snapshots, and $W$ is the complexity of each snapshot submodel.

\section{Additional Results}
\label{sec:additional_results}

\textbf{Zero-Shot Performance}
Table~\ref{tab:zero_shot_F1} presents the zero-shot performance of CodeLlama in two popular code analysis tasks. 
We observe a noticeable challenge: the model produces nearly-zero F-1 in all datasets.
This suggests current capabilities of CodeLlama in zero-shot settings, especially when deployed for solving complex tasks, are still inadequate, underscoring further fine-tuning in real-world code analysis applications.

\begin{figure*}[]
\centering
    \begin{tabular}{c c}
    \rotatebox[origin=c]{90}{\small{(a)~CC}} & \includegraphics[width=0.96\textwidth,valign=m]{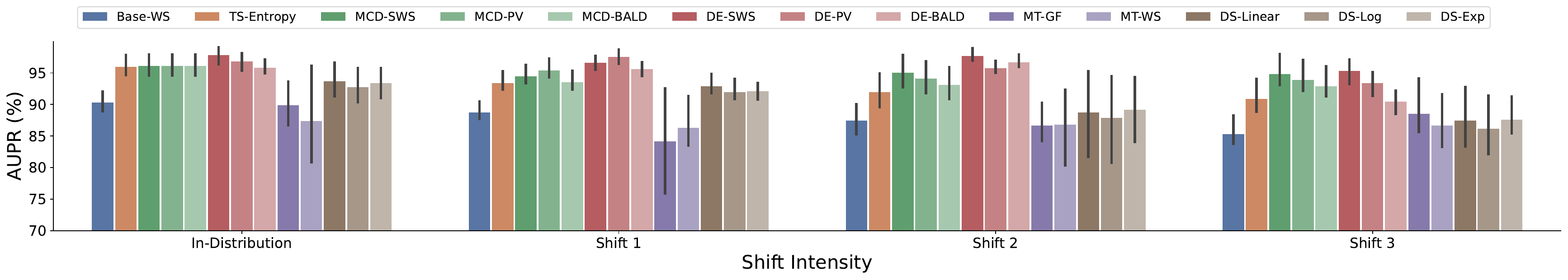}
    \\
    \rotatebox[origin=c]{90}{\small{(b)~CS}} & \includegraphics[width=0.96\textwidth,valign=m]{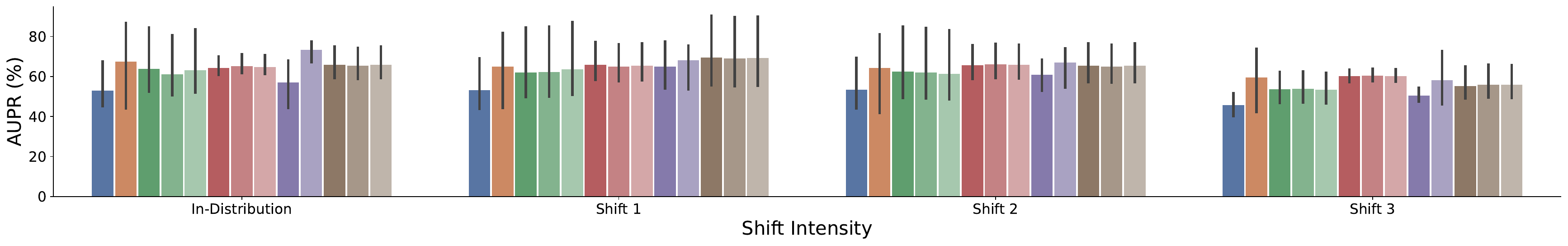}
    \end{tabular}
    \caption{Misclassification results (AUPR) for probabilistic methods in (a) CC and (b) CS, with intensity shifts. Error bars indicate AUPR variations across the three shift patterns.}
    \label{fig:mis_detect_aupr}
\end{figure*}

\begin{figure*}
\centering
\includegraphics[width=0.98\textwidth]{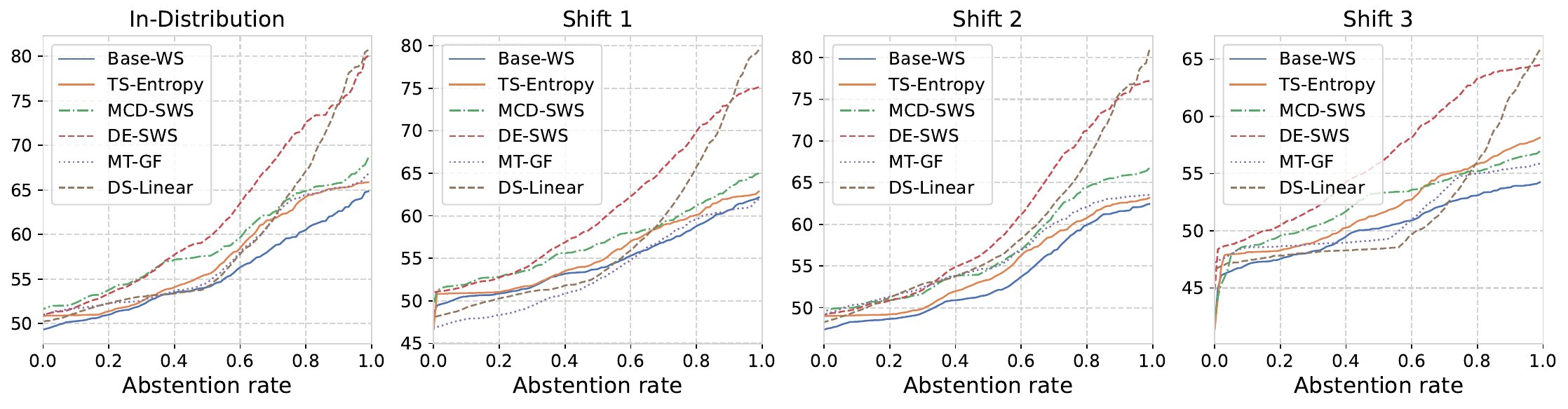}
\caption{F-1 (\%) vs. Abstention curve for different uncertainty methods in CS under distribution shifts. Each line is the average F-1 score over the three shift patterns.}
\label{fig:filter_cs}
\end{figure*}

\begin{figure*}[]
\centering
    \begin{tabular}{c c}
    \rotatebox[origin=c]{90}{\small{(a)~Code completion}} & \includegraphics[width=0.96\textwidth,valign=m]{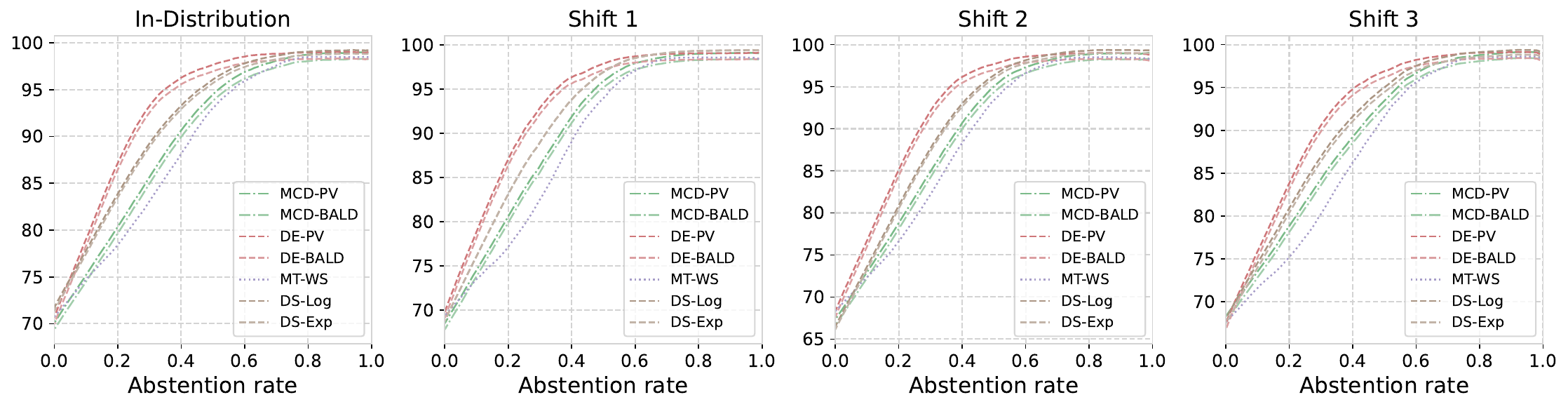}
    \\
    \rotatebox[origin=c]{90}{\small{(b)~Code summarization}} & \includegraphics[width=0.96\textwidth,valign=m]{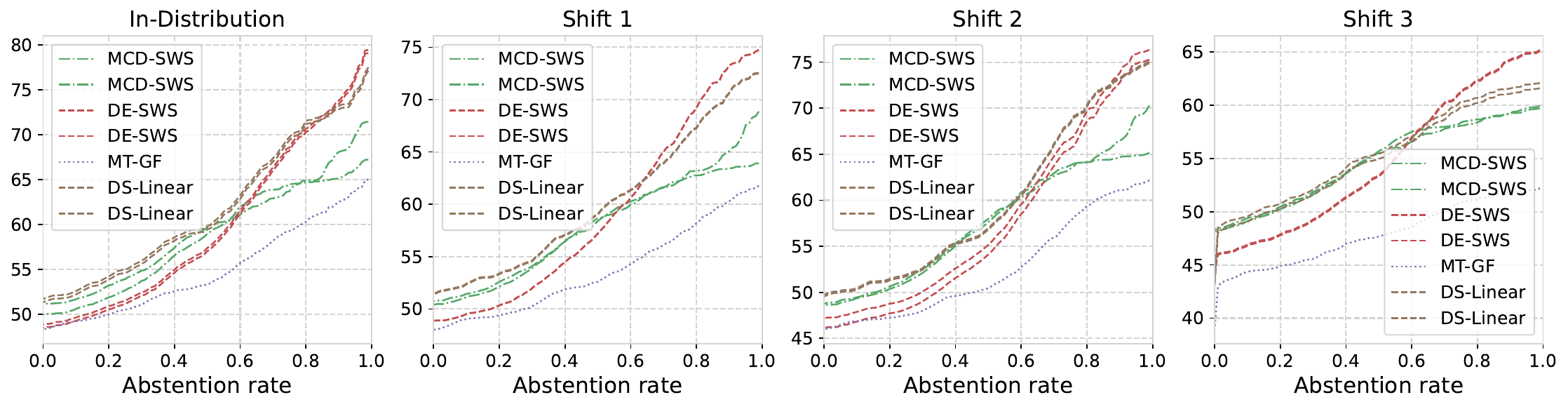}
    \end{tabular}
    \caption{F-1 (\%) vs. Abstention curve for the remaining UE methods in both in-distribution and shifted code examples. Each line is the average F-1 score over the three shift patterns.}
    \label{fig:filter_remain}
\end{figure*}


\begin{figure*}
\centering
   \begin{tabular}{c c}
    \rotatebox[origin=c]{90}{\small{(a)~Code completion}} & \includegraphics[width=0.65\textwidth,valign=m]{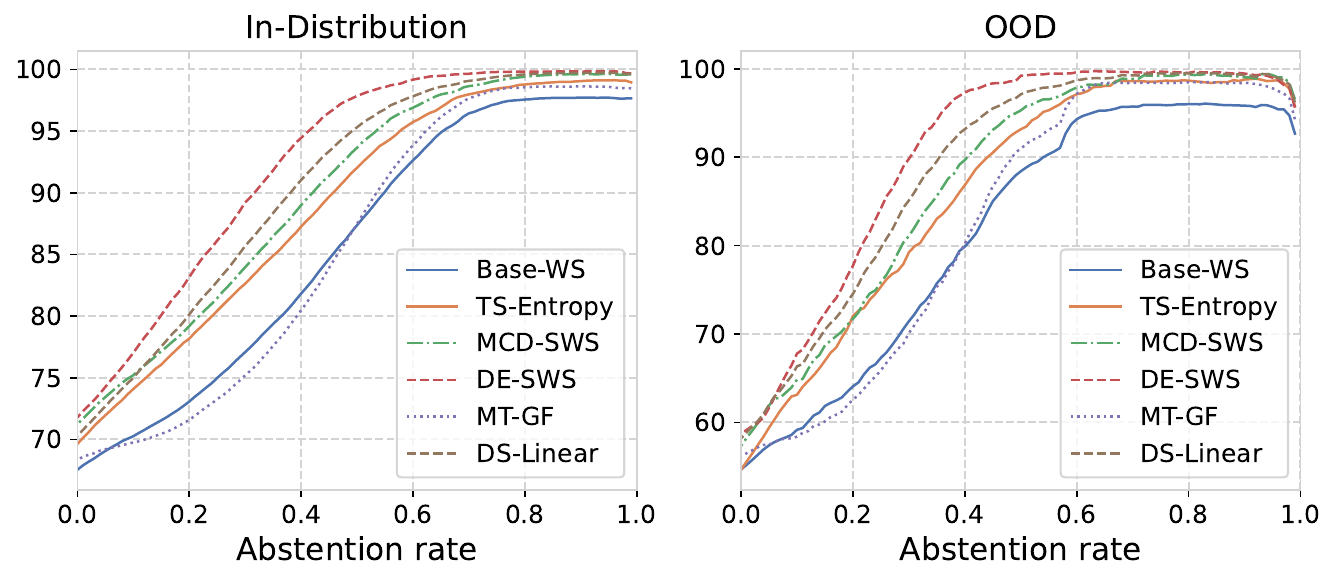}
    \\
    \rotatebox[origin=c]{90}{\small{(b)~Code summarization}} & \includegraphics[width=0.65\textwidth,valign=m]{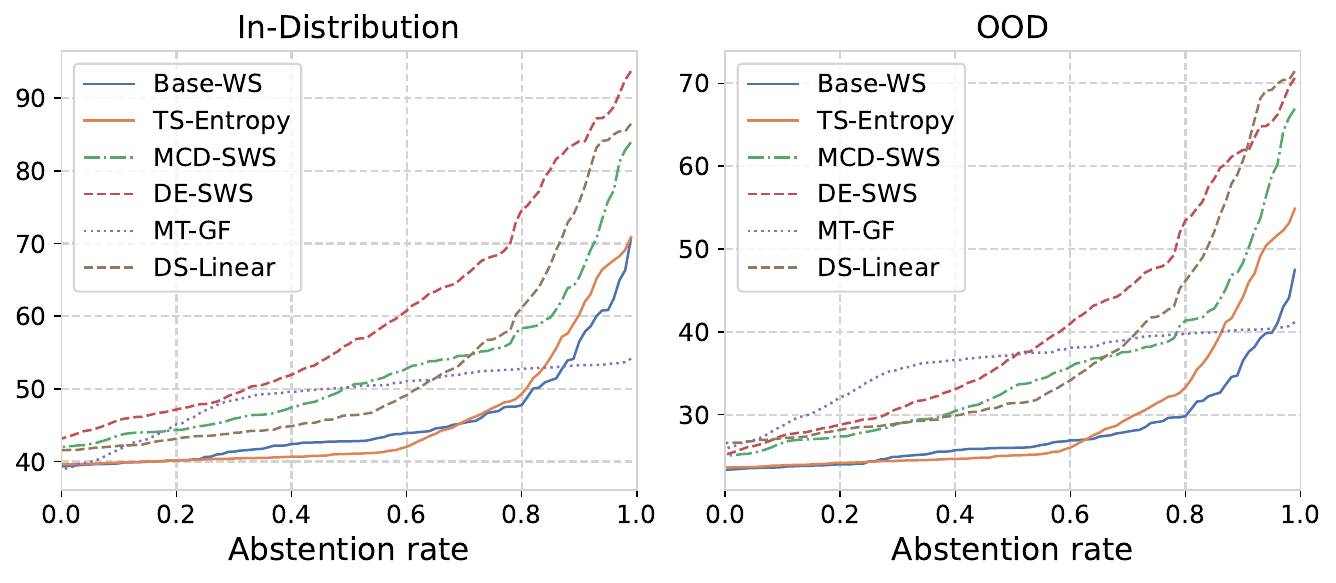}
    \end{tabular}
   \caption{F-1 (\%) vs. Abstention curve for different uncertainty methods in both in-distribution and OOD code samples.}
   \label{fig:input_val_ood}
\end{figure*}

\textbf{Misclassification Detection}
Figure~\ref{fig:mis_detect_cs} shows the AUC and Brier results of misclassification detection for different methods in CS. 
We observe a similar pattern as in CC: DS is superior in validating low-quality predictions or mistakes.
Figure~\ref{fig:mis_detect_aupr} presents the AUPR results of misclassification detection under the three shifted datasets.
Consistent with the AUC and Brier performance, we also observe a degradation of quality of UE methods under distribution shifts, as evidenced by lower AUPR on the shift3 set.
Probabilistic methods generally improves the mistakes awareness over the vanilla baseline.
DE, MCD, and DS achieve promising AUPR results.
Interestingly, post-hoc calibration (TS) is prone to degradation under intensifying shifts.


\textbf{Selective Prediction}
Figure~\ref{fig:filter_cs} demonstrates the selective prediction results in CS. We observe probabilistic methods generally lead to higher F-1 performance with the same rejection rate compared to the vanilla baseline. 
Among them, DE and DS still achieve higher efficiency.
However, compared to the results in CC, the highest F-1 is much smaller and cannot reach near 100\%. This is because the prediction quality of the vanilla model in CS is much lower than in CC, causing the UE a bit more challenge.
Figure~\ref{fig:filter_remain} illustrates the selective prediction results of the remaining UE methods. We observe different UE techniques that corresponds to the same probabilistic method produce similar results. DE and DS are efficient in achieving the same prediction quality with less abstention rate.
Figure~\ref{fig:input_val_ood} demonstrates the abstention prediction performance for both in-distribution and OOD examples.
We observe a more severe degradation of efficiency.
Among them, DE and DS still achieve higher F-1 scores in OOD dataset, while the MT is relatively less effective compared to other probabilistic methods.
This again suggests its UE assumption is sensitive to severe shift but less relevant to the reliability of model predictions.

\end{document}